\documentclass{article}
\usepackage[utf8]{inputenc}
\usepackage{amsfonts}
\usepackage{mathrsfs,amsmath,latexsym}
\usepackage{float}
\usepackage{hyperref,graphicx}
\usepackage{geometry}
\usepackage{braket}
\usepackage{physics}
\usepackage[parfill]{parskip}
\usepackage{multicol}
\usepackage{color,soul}
\usepackage{chemformula}
\usepackage{wasysym}
\usepackage{caption}

\usepackage[utf8]{inputenc}
\usepackage[sorting=none,citestyle=numeric-comp, url=false, eprint=false, isbn=false, backend=bibtex, maxnames=3, doi=true]{biblatex}
\AtEveryBibitem{\clearfield{month}}
\usepackage{authblk}

\newcommand\tinyhexagon{\vcenter{\hbox{\scalebox{0.8}{$\hexagon$}}}}
\newcommand\tinyhexagonn{\vcenter{\hbox{\scalebox{0.6}{$\hexagon$}}}}

\title{Ultrafast coherent magnon spin currents in antiferromagnets}
\author[1]{Torstein Hegstad}
\author[1]{Johan H. Mentink}
\affil[1]{Radboud University, Institute for Molecules and Materials (IMM) Heyendaalseweg 135, 6525 AJ Nijmegen, The Netherlands}

\begin{document}
\maketitle

%\section{Abstract}
% \twocolumn[
%   \begin{@twocolumnfalse}
%     \maketitle
%     \begin{abstract}
%     \noindent Generating coherent magnon spin currents with the highest frequencies and shortest wavelengths is a key challenge in ultrafast spintronics and magnonics. A promising route is to excite counter-propagating magnon pairs. In antiferromagnets, such pairs can be accessed in the ultrafast regime, where coherent dynamics are dominated by magnons at the edge of the Brillouin zone. However, it has seemed impossible to generate a net spin current from coherent magnon pairs. Here we show that a coherent superposition of multiple magnon-pair modes can produce such a current in parity-time symmetric antiferromagnets. 
%     The ultrafast coherent spin currents are excited with linearly polarized light, with the light polarization steering the current direction.
%     Finally, by superposing two orthogonal spin currents, circular spin currents can be generated, which have not been discussed for steady-state currents.
%             %Modelling a prototypical 2D honeycomb antiferromagnet we show that, unlike steady-state currents, these ultrafast spin currents can be driven by linearly polarized light, with polarization steering their direction. 
%     \end{abstract}
%     \vspace{1em}
%   \end{@twocolumnfalse}
%]

\begin{abstract}
\noindent Generating coherent magnon spin currents with the highest frequencies and shortest wavelengths is a key challenge in ultrafast spintronics and magnonics. A promising route is to excite counter-propagating magnon pairs. In antiferromagnets, such pairs can be accessed in the ultrafast regime, where coherent dynamics are dominated by magnons at the edge of the Brillouin zone. However, it has seemed impossible to generate a net spin current from coherent magnon pairs. Here we show that a coherent superposition of multiple magnon-pair modes can produce such a current in parity-time symmetric antiferromagnets. 
The ultrafast coherent spin currents are excited with linearly polarized light, with the light polarization steering the current direction.
Finally, by superposing two orthogonal spin currents, circular spin currents can be generated, which have not been discussed for steady-state currents.
\end{abstract}
\begin{multicols}{2}

%\section{Introduction}

Transporting information using the spin degree of freedom plays a central role in the field of spintronics and magnonics \cite{grundlerNanomagnonicsCorner2016,pirroAdvancesCoherentMagnonics2021,zeleznySpinTransportSpin2018, hanCoherentAntiferromagneticSpintronics2023}. Magnons, the collective excitations of magnetically order systems, are especially interesting for this purpose, since they can carry a net spin even without requiring motion of the electron charge. Therefore, as opposed to electronic spin currents, magnons do not experience Joule heating, exhibit inherently low damping, and are therefore extensively studied for their potential to realize energy-efficient spin transport \cite{chumakMagnonSpintronics2015,flebus2024MagnonicsRoadmap2024,flebusRecentAdvancesMagnonics2023, gillPureSpinCurrents2025, gillUltrafastAllopticalGeneration2025}. However, distinct from excitation of electron-spin current, for which many proposals to excite them on femtosecond time scales exists \cite{gillUltrafastAllopticalGeneration2025, guptaTuningUltrafastDemagnetization2025,sharmaTHzInducedGiant2023,maPhotocurrentMultiphysicsDiagnostic2023}, generation of magnon spin currents on femtosecond time scales remains a major challenge.

To address this challenge, the class of antiferromagnets (AFM) materials are of great interest, as their magnons modes are in the THz range as compared to GHz for ferromagnets. Although recently great progress has been made to excite propagating antiferromagnetic magnons wave packets \cite{hortensiusCoherentSpinwaveTransport2021} with femtosecond optical pulses, it relies on confining the excitation volume. The magnon wavelengths thus generated will still be close to the center of the Brillouin zone (BZ). Much higher frequencies can be generated by exciting magnons at the edge of the BZ. In principle, such magnons can be excited with optical pulses as well, by exciting them as a pair of counter-propagating magnons pairs by perturbing the exchange interaction, such that the net transferred momentum remains zero \cite{fleuryScatteringLightOne1968, fedianinSelectionRulesUltrafast2023}.

Interestingly, such two-magnon modes (2M) have been studied experimentally in the coherent regime \cite{formisanoCoherentTHzSpin2024, bossiniLaserdrivenQuantumMagnonics2019, zhaoMagnonSqueezingAntiferromagnet2004} and theoretical predictions have shown that the excitation of two-magnons wave packets can lead to propagation of spin correlations at supermagnonic velocities \cite{fabianiSupermagnonicPropagationTwodimensional2021}. However, despite the fact that proposals exists for optical generation of magnon spin currents under continous illumination \cite{bostromAllopticalGenerationAntiferromagnetic2021,fujiwaraNonlinearSpinCurrent2023, proskurinExcitationMagnonSpin2018, ishizukaLargePhotogalvanicSpin2022}, it remains an open problem whether it is feasible at all to generate a net magnon spin current by coherently excited magnon pairs with ultrashort optical pulses. In particular, unlike the generation of spin currents by coherent single-magnon excitation, the net spin angular momentum of magnon pairs is zero.

In this paper we show that ultrafast generation of a coherent 2M spin current should be possible in a parity-time symmetric antiferromagnet (PT AFM), which are currently widely studied for their relevance in 2D van der Waals systems \cite{allingtonDistinctOpticalExcitation2025, freemanTunableUltrastrongMagnon2025,kurebayashiMagnetismSymmetrySpin2022,afanasievControllingAnisotropyVan2021,khanRecentBreakthroughsTwodimensional2020,wangProspectsOpportunities2D2020, yangVanWaalsMagnets2021}. Moreover, we argue that the effect also is possible in AFMs with a lower symmetry. The spin current emerges even with linear polarized light and utilizes the inherent parity symmetry breaking of the underlying lattice of the spins. Moreover, the spin current direction is controllable with the light polarization, and due to the coherent nature, a double-pulse excitation even allows for generation of a circularly polarized magnons spin current, which has neither been predicted nor measured before.

\begin{figure}[H]
    \centering
    \includegraphics[scale=0.6]{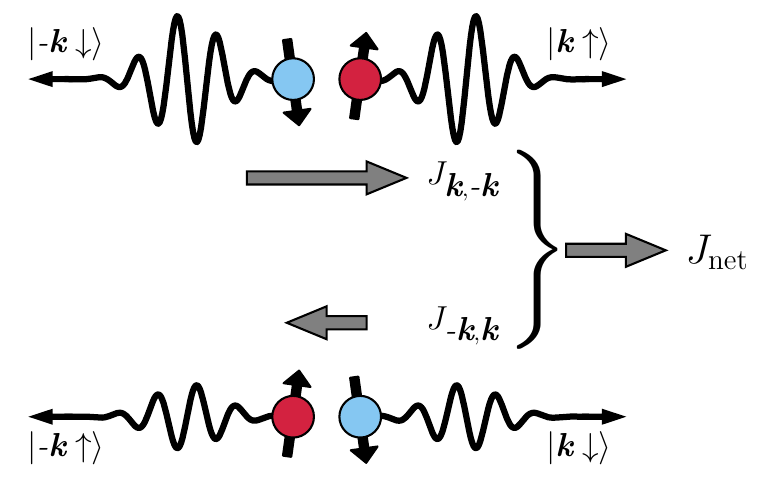}
    \caption{Schematic picture two 2M modes propagating in opposite directions with different probability amplitude. Each of the 2M modes is responsible for the currents components $J_{\vb*{k},-\vb*{k}}$ and $J_{-\vb*{k},\vb*{k}}$ respectively, which in this schematic do not cancel, and results in a net current $J_{\mathrm{net}}$}
    \label{fig: 2Mcartoon_current}
\end{figure}

The key mechanism is illustrated in Figure \ref{fig: 2Mcartoon_current}. In conventional AFMs, the impulsive excitation of 2M pairs comprises a coherent superposition of many magnon pairs. Although a single pair $\ket{\vb*{k}\uparrow, -\vb*{k}\downarrow}$, illustrated at the top of the figure, will give rise to a spin current $J_{\vb*{k},-\vb*{k}}$, this current is exactly cancelled the pair $\ket{-\vb*{k}\uparrow, \vb*{k}\downarrow}$, which by symmetry must also be excited and gives the same yet opposite spin current $J_{-\vb*{k}, \vb*{k}}$. Instead, if parity-symmetry is broken, the magnon pairs $\ket{\vb*{k}\uparrow, -\vb*{k}\downarrow}$ and $\ket{-\vb*{k}\uparrow, \vb*{k}\downarrow}$ attain different phases and do not deconstructivly interfere with each other. Thus, a net spin current can emerge. It is important to note that even though parity symmetry is broken, Dzyaloshinskii-Moriya interaction is not required for this effect. Moreover, the total spin angular momentum remains zero in the system.

% \begin{center}
%   \includegraphics[scale=0.6]{figures/2Mcartoon_current.pdf}
%   \captionof{figure}{Schematic picture two 2M modes propagating in opposite directions with different probability amplitude. Each of the 2M modes is responsible for the currents components $J_{\vb*{k},-\vb*{k}}$ and $J_{-\vb*{k},\vb*{k}}$ respectively, which in this schematic do not cancel, and results in a net current $J_{\mathrm{net}}$}
%   \label{fig: 2Mcartoon_current}
% \end{center}

% \subsection{Paper structure}

% The remainder of the paper is structured as follows. We start by introducing the model of the AFM and the light matter-interaction. We then solve it with LSWT \hl{First time?}, and use it to calculate a magnon spin current. Next, we show coherent control of the current by predicting a circular current. Lastly there is a discussion and conclusion.

% \subsection{Articles to cite}. This on is about ultrafast spin current in antiferromagnets: \cite{qiuUltrafastSpinCurrent2021}.

%\section{Model}

To illustrate the effect, we focus on the nearest neighbor Heisenberg model on the 2D honeycomb lattice, which is a minimal model of a PT AFM. We model the excitation of the magnon pairs using the generic model for perturbation of exchange interactions \cite{mentinkUltrafastReversibleControl2015,losadaUltrafastControlSpin2019}, consistent with 2M Raman scattering \cite{fleuryScatteringLightOne1968, fedianinSelectionRulesUltrafast2023, formisanoCoherentTHzSpin2024, bossiniLaserdrivenQuantumMagnonics2019,zhaoMagnonSqueezingAntiferromagnet2004,bossiniMacrospinDynamicsAntiferromagnets2016}. The total Hamiltonian can then be written as
\begin{equation}
    \label{eq: Heisenberg hamiltonian}
    \mathcal{H} = \mathcal{H}_0 + \delta\mathcal{H} = \sum_{i, \vb*{\delta}} \left[J_0 +  \Delta J f(t) (\vb*{e}\cdot \vb*{\delta})^2 \right] \vb*{S}_i \cdot \vb*{S}_{i+\delta}
\end{equation}
where $\vb*{S}_i = \vb*{S}(\vb*{r}_i)$ is the spin operator on lattice site $i$, and $\vb*{S}_{i+\delta} = \vb*{S}(\vb*{r}_i + \vb*{\delta})$ is the spin operator on the nearest neighbor of $\vb*{S}_i$ defined by the lattice vector $\vb*{\delta}$. $J_0$ is the nearest neighbor exchange constant, $\Delta J$ is the exchange perturbation strength, and $\vb*{e} = (\cos\theta, \sin\theta)$ is the polarization of the electric field of the incoming radiation with angle $\theta$ from the $x$-axis, which propagates normal to the plane of the magnetic medium. This means that the light only perturbs bonds $\vb*{\delta}$ with a projection along $\vb*{e}$. The shape of the pulse is given by $f(t)=\tau\delta(t)$, where $\tau$ is the duration of the pulse, and $\delta(t)$ is the Dirac distribution function. Unless otherwise specified, we set $\hbar=1$ and the lattice constant $a=1$ in this work.

To solve for the dynamics, we employ linear spin-wave theory. In PT antiferromagnets, nearest-neighbor spins are not equivalent, and we thus introduce two distinct bosonic operators $a_i$ and $b_i$. We do this by rewriting the unperturbed Hamiltonian, $\mathcal{H}_0$, in Eq. \eqref{eq: Heisenberg hamiltonian} using the Holstein-Primakoff (HP) transformation with $S^z_{iA} = S-a^\dag_i a_i$, $S_{iA}^+ = \sqrt{2S}a_i$, $S_{iA}^- = \sqrt{2S}a_i^\dag$ on sublattice A; while for the B sublattice we have $S^z_{iB} = -S+b^\dag_i b_i$, $S_{iB}^+ = \sqrt{2S}b_i^\dag$, and $S_{iB}^- = \sqrt{2S}b_i$. $S$ is the spin value, and $S_i^{\pm} = S_i^x \pm i S_i^y$. Since the effect we aim to simulate derives directly from symmetry, it is sufficient to limit ourselves to linear spin-wave theory (LSWT) in which only terms up to second order in the boson operators are taken into account. For the description of spin waves, it is convenient to work in reciprocal space using the Fourier transforms given by  $a_i = \sqrt{2/N} \sum_{\vb*{k}} e^{i \vb*{k}\cdot \vb*{r}_i} a_{\vb*{k}}$ and $b_i = \sqrt{2/N} \sum_{\vb*{k}} e^{i \vb*{k}\cdot \vb*{r}_i} b_{\vb*{k}}$. Apart from a constant term, we obtain
\begin{equation}
    \label{eq: HP Ham}
    \mathcal{H}_0 = zJ_0 S  \sum_{\vb*{k}} \begin{bmatrix}
        a^\dag_{\vb*{k}} & b_{-\vb*{k}}
    \end{bmatrix}
    \begin{bmatrix}
        1 & \gamma_{\vb*{k}} \\
        \gamma_{\vb*{k}}^* & 1
    \end{bmatrix}
    \begin{bmatrix}
        a_{\vb*{k}} \\ b_{-\vb*{k}}^\dag
    \end{bmatrix},
\end{equation}
where $z$ is the number of nearest neighbors and $N$ is the number of spins in the system. The factor $\gamma_{\vb*{k}} =z^{-1}\sum_{\vb*{\delta}} e^{i\vb*{k}\cdot\vb*{\delta}} = |\gamma_{\vb*{k}}|e^{i\varphi_{\vb*{k}}}$ encodes the lattice structure and some of the symmetries of the system. E.g. the phase factor $\varphi_{\vb*{k}}$ only appears in parity-breaking systems. We diagonalize Eq. \eqref{eq: HP Ham} by applying the Bogoliubov transformation
\begin{equation}
    \label{eq: Bogoliubov transformation}
    \begin{bmatrix}
        a_{\vb*{k}} \\
        b_{-\vb*{k}}^\dag
    \end{bmatrix}
    =
    \begin{bmatrix}
        u_{\vb*{k}} e^{i\varphi_{\vb*{k}}} & v_{\vb*{k}} e^{i\varphi_{\vb*{k}}}\\
        v_{\vb*{k}} & u_{\vb*{k}}
    \end{bmatrix}
    \begin{bmatrix}
        \alpha_{\vb*{k}} \\
        \beta_{-\vb*{k}}^\dag
    \end{bmatrix},
\end{equation}
where $\alpha_{\vb*{k}}$ and $\beta_{\vb*{k}}$ are the boson operators in the transformed frame. Inserting Eq. \eqref{eq: Bogoliubov transformation} into Eq. \eqref{eq: HP Ham} we get
\begin{equation}
    \mathcal{H} = \sum_{\vb*{k}} \omega_{\vb*{k}} \left[ \alpha_{\vb*{k}}^\dag \alpha_{\vb*{k}} + \beta_{-\vb*{k}}^\dag \beta_{-\vb*{k}} \right],
\end{equation}
where the energy dispersion relation is given by $\omega_{\vb*{k}} = zJ_0 S \epsilon_{\vb*{k}}$, where $\epsilon_{\vb*{k}} = \sqrt{1-|\gamma_{\vb*{k}}|^2}$. Using the same transformations on $\delta H$ in Eq. \eqref{eq: Heisenberg hamiltonian} yields
\begin{equation}
    \delta\mathcal{H} = f(t) \sum_{\vb*{k}}
    \begin{bmatrix}
        \alpha_{\vb*{k}}^\dag &
        \beta_{-\vb*{k}}
    \end{bmatrix}
    \begin{bmatrix}
        \delta \omega_{\vb*{k}} & V_{\vb*{k}}\\
        V_{\vb*{k}}^* & \delta \omega_{\vb*{k}}
    \end{bmatrix}
    \begin{bmatrix}
        \alpha_{\vb*{k}} \\
        \beta_{-\vb*{k}}^\dag
    \end{bmatrix}.
\end{equation}
$\delta \omega_{\vb*{k}}$ is the perturbation to the dispersion relation, and
% \begin{equation}
%     V_{\vb*{k}} =\Delta J S\left( \frac{1}{\epsilon_{\vb*{k}}}\real\left\{ \xi_{\vb*{k}} e^{-i\varphi_{\vb*{k}}} \right\} + i \imaginary\left\{ \xi_{\vb*{k}} e^{-i\varphi_{\vb*{k}}} \right\} - \frac{\zeta |\gamma_{\vb*{k}}|}{\epsilon_{\vb*{k}}} \right)
% \end{equation}
\begin{equation}
    V_{\vb*{k}} =\Delta J S\left( \frac{\mathrm{Re}\left\{ \xi_{\vb*{k}} e^{-i\varphi_{\vb*{k}}} \right\} - \zeta |\gamma_{\vb*{k}}|}{\epsilon_{\vb*{k}}} + i \mathrm{Im}\left\{ \xi_{\vb*{k}} e^{-i\varphi_{\vb*{k}}} \right\}\right),
\end{equation}
where $\xi_{\vb*{k}} = \sum_\delta (\vb*{e}\cdot \vb*{\delta})^2 e^{i{\vb*{k}}\cdot \vb*{\delta}}$ and $\zeta = \sum_{\vb*{\delta}} (\vb*{e}\cdot\vb*{\delta})^2=3/2$ for the honeycomb lattice. Within linear response, the dynamics can be solved analytically (see supplementary Sec. S.1 for derivation)\cite{boumanTimedependentSchwingerBoson2024}:
\begin{equation}
    \langle \alpha_{\vb*{k}}^\dag \beta_{-\vb*{k}}^\dag \rangle = V_{\vb*{k}}^* \tau i e^{2i\omega_{\vb*{k}}t}.
\end{equation}
and $\langle \alpha_{\vb*{k}} \beta_{-\vb*{k}} \rangle$ is its complex conjugate. The magnon densities $\langle \alpha_{\vb*{k}}^\dag \alpha_{\vb*{k}} \rangle = \langle \beta_{-\vb*{k}}^\dag \beta_{-\vb*{k}} \rangle$ contributes only in the second order $(\Delta J)^2$, which we ignore in the calculations below. Moreover, they are even functions in momentum $\vb*{k}$ and hence are not subject to the effect originating from parity symmetry breaking that has our interest. These results give the well-known dynamics of spin correlations previously investigated \cite{fedianinSelectionRulesUltrafast2023, boumanTimedependentSchwingerBoson2024, fabianiParametricallyDrivenTHz2022}

Next, we define the magnon spin current. The $z$ component of the total spin is conserved, thus the local magnon density $n(\vb*{r}_i) = \sum_{\vb*{\delta}} [b_{i+\delta}^\dag b_{i+\delta}- a_i^\dag a_i]$ should satisfy a continuity equation, which can be expressed as
$\partial_t n_{\vb*{q}}+ i\vb*{q}\cdot \vb*{J}=0$ as $\vb*{q} \to 0$ \cite{proskurinExcitationMagnonSpin2018}. The current is then given by[ref, appendix]
\begin{equation}
    \label{eq: current operator}
    \vb*{J} = \frac{2}{N} \sum_{\vb*{k}} \begin{bmatrix}
        \alpha^\dag_{\vb*{k}} & \beta_{-\vb*{k}}
    \end{bmatrix}
    \begin{bmatrix}
        \nabla_{\vb*{k}} \omega_{\vb*{k}} & \vb*{K}_{\vb*{k}} \\
        \vb*{K}_{\vb*{k}}^* & \nabla_{\vb*{k}} \omega_{\vb*{k}}
    \end{bmatrix}
    \begin{bmatrix}
        \alpha_{\vb*{k}} \\ \beta_{-\vb*{k}}^\dag
    \end{bmatrix},
\end{equation}
where
% \begin{equation}
%     \vb*{K}_{\vb*{k}} =zJ_0S \left[ \frac{1}{\epsilon_{\vb*{k}}} \nabla_{\vb*{k}} |\gamma_{\vb*{k}}| + i |\gamma_{\vb*{k}}| \nabla_{\vb*{k}} \varphi_{\vb*{k}} \right]
% \end{equation}
\begin{equation}
    \vb*{K}_{\vb*{k}} =zJ_0S \left[  i |\gamma_{\vb*{k}}| \nabla_{\vb*{k}} \varphi_{\vb*{k}} -  \frac{\nabla_{\vb*{k}} \epsilon_{\vb*{k}}}{|\gamma_{\vb*{k}}|}   \right].
\end{equation}

Previous works \cite{bostromAllopticalGenerationAntiferromagnetic2021, proskurinExcitationMagnonSpin2018}, focusing on the steady state DC regime, mainly consider the diagonal terms in Eq.~\eqref{eq: current operator}. In contrast, since the group velocity $\nabla_{\vb*{k}}\omega_{\vb*{k}}$ is an odd function in $\vb*{k}$, the diagonal terms vanish as we sum over the BZ. Instead, the off diagonal terms, together with the expectation value of the two magnon modes, contain even terms and can hence contribute to a net current. The expectation value of the current can be written as $\langle \vb*{J} \rangle~=~\frac{4}{N}\sum_{\vb*{k}} \mathrm{Re} \{ \vb*{K}_{\vb*{k}} \langle \alpha_{\vb*{k}}^\dag \beta_{-\vb*{k}}^\dag \rangle \} = \sum_{\vb*{k}} \vb*{J}_{\vb*{k}}$ with
\begin{equation}
    \label{eq: Jk general}
    \vb*{J}_{\vb*{k}} = \Gamma\bigg[ |\gamma_{\vb*{k}}| \nabla_{\vb*{k}} \varphi_{\vb*{k}} \mathrm{Re}\{V_{\vb*{k}}\} + \frac{\nabla_{\vb*{k}}\epsilon_{\vb*{k}}}{|\gamma_{\vb*{k}}|} \mathrm{Im}\{V_{\vb*{k}}\}\bigg] \cos(2\omega_{\vb*{k}}t),
\end{equation}
and $\Gamma = -4zJ_0 S \tau /N$. Terms odd in $\vb*{k}$ have been excluded.
Note that the effect is nonlinear in the electric field, since the perturbation $\delta\mathcal{H}\sim E^2$, despite that all calculations are obtained in linear response for the products of magnon operators. The expression in Eq.~\eqref{eq: Jk general} consists of the two terms. The first arises from the lack of parity symmetry in the system, and $\nabla_{\vb*{k}} \varphi_{\vb*{k}}$ is closely related to the Berry connection and Berry curvature. In the case of parity symmetry, this term vanishes. The second is like the diagonal part of the current, proportional to the group velocity. However, $\mathrm{Im}\{V_{\vb*{k}}\}$ represents the lack of parity symmetry in the perturbation, and is odd in $\vb*{k}$. Thus, this term also gives a non-zero contribution to the current. Note that the discussion above focuses entirely on the parity symmetry breaking. Using LSWT, time-reversal symmetry breaking is assumed throughout. If we would construct a modified LSWT in which time-reversal symmetry is not broken, we would get an additional sum over all spin states with sign corresponding to the $z$-component giving zero net current.

So far the expressions obtained hold for general Heisenberg AFMs. For the evaluation of the spin current below we focus on the 2D honeycomb lattice, which is currently widely investigated in spintronics and magnonics research \cite{goMagnonOrbitalNernst2024,chenDampedTopologicalMagnons2023,yangVanWaalsMagnets2021,losadaUltrafastControlSpin2019}. In this case, the expressions can be further simplified using the specific expression for $V_{\vb*{k}}$ giving
% \begin{equation}
%     \begin{split}
%         \langle J_\mu \rangle =& \frac{18}{N}\Delta J J_0 S^2 \tau \Theta_\mu \times\\
%         &\sum_{\vb*{k}} \frac{1}{\epsilon_{\vb*{k}}} \bigg[ (|\gamma_{\vb*{k}}| \partial_{k_\mu} \varphi_{\vb*{k}})^2 + (\partial_{k_{\mu}}|\gamma_{\vb*{k}}|)^2 \bigg] \cos(2\omega_{\vb*{k}}t)
%     \end{split}
%     \label{eq: current honeycomb org}
% \end{equation}
% \begin{equation}
%     \langle J_\mu \rangle = \frac{18}{N}\Delta J J_0 S^2 \tau \Theta_\mu \sum_{\vb*{k}} \frac{|\partial_{k_\mu}\gamma_{\vb*{k}}|^2}{\epsilon_{\vb*{k}}}  \cos(2\omega_{\vb*{k}}t)
%     \label{eq: current honeycomb}
% \end{equation}
\begin{equation}
    \label{eq: current honeycomb}
    \langle J_\mu \rangle = \Lambda\Theta_\mu
    \sum_{\vb*{k}} \bigg[ (|\gamma_{\vb*{k}}| \partial_{k_\mu} \varphi_{\vb*{k}})^2 + (\partial_{k_{\mu}}|\gamma_{\vb*{k}}|)^2 \bigg]  \frac{\cos(2\omega_{\vb*{k}}t)}{\epsilon_{\vb*{k}}},
\end{equation}
% \hl{or?}
% \begin{equation}
%     \label{eq: current honeycomb}
%     \langle J_\mu \rangle = \Lambda\Theta_\mu
%     \sum_{\vb*{k}} \bigg[ \left( \frac{|\gamma_{\vb*{k}}| \partial_{k_\mu} \varphi_{\vb*{k}}}{\epsilon_{\vb*{k}}}\right)^2 + \left(\frac{\partial_{k_{\mu}}\epsilon_{\vb*{k}}}{|\gamma_{\vb*{k}}|} \right)^2 \bigg]\cos(2\omega_{\vb*{k}}t)
% \end{equation}
where $\vb*{\Theta}=(\sin2\theta, \cos2\theta)$, $\Lambda=\frac{18}{N}\Delta J J_0 S^2 \tau$, and $\mu\in \{x,y\}$. We have used the lattice vectors given by $\vb*{\delta}_1 = \frac{1}{2} (\sqrt{3}, -1), \vb*{\delta}_2 = (0,1)$, and $\vb*{\delta}_3 = -\frac{1}{2}(\sqrt{3}, 1)$. The polarization dependence of $2\theta$ originates from the quadratic dependence of the electric field, and lattice symmetries. Note that an extra factor of $1/a$ is needed to give Eq. \eqref{eq: current honeycomb} the correct units.

%\section{Results and discussion}

Investigating Eq.\eqref{eq: current honeycomb} we find a polarization dependent selection rule in the case of a honeycomb lattice. The sum over $\vb*{k}$ yield the same for both components of the current. Hence, the current can be written as $\langle \vb*{J} \rangle = A(t) \vb*{\Theta}$, where $A(t)$ is a time dependent amplitude. The direction of the current for the honeycomb lattice is thus dependent on two times the polarization direction; $2\theta$.

\begin{figure}[H]
    \centering
    \includegraphics[scale=0.55]{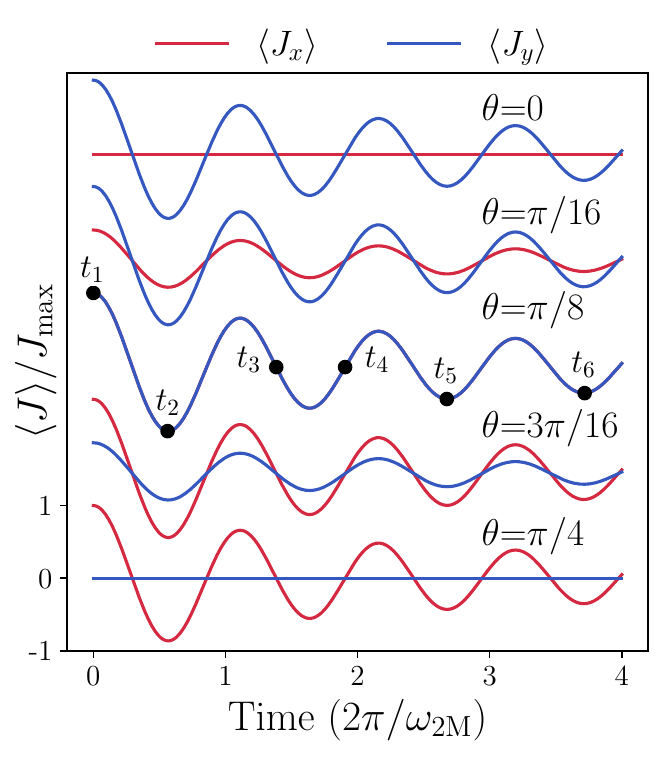}
    \caption{Current as a function of time for different polarization $\theta$. The times $t_1 \ldots t_6$ represents snapshots in time used later in Figure \ref{fig: Jk snapshots}.}
    \label{fig: basic current}
\end{figure}

Figure \ref{fig: basic current} shows the time-dependence of the spin current as a function of polarization angle in accordance with Eq.~\eqref{eq: current honeycomb}. The amplitude of the current is normalized with $J_{\mathrm{max}} = |\langle \vb*{J} \rangle_{t=0}|$ (magnitude discussed in more detail later).

% \begin{center}
%   \includegraphics[scale=0.55]{figures/polarization_waterfall.pdf}
%   \captionof{figure}{Current as a function of time for different polarization $\theta$. The times $t_1 \ldots t_6$ represents snapshots in time used later in Figure \ref{fig: Jk snapshots}.}
%   \label{fig: basic current}
% \end{center}

The oscillation periods are close to the two-magnon frequency at the edge of the BZ, $\omega_{\mathrm{2M}} = 2zJ_0 S$, while the amplitude decreases as a function of time. This is consistent with the dominant contribution to the current stemming from the magnon pairs at the edge of the BZ, while the decrease of the amplitude stems from dephasing of magnon pairs with different frequency, which also causes a slightly time-dependent oscillation frequency, as we further explain below.

\begin{figure*}[t]
    \centering
    \includegraphics[scale=0.53]{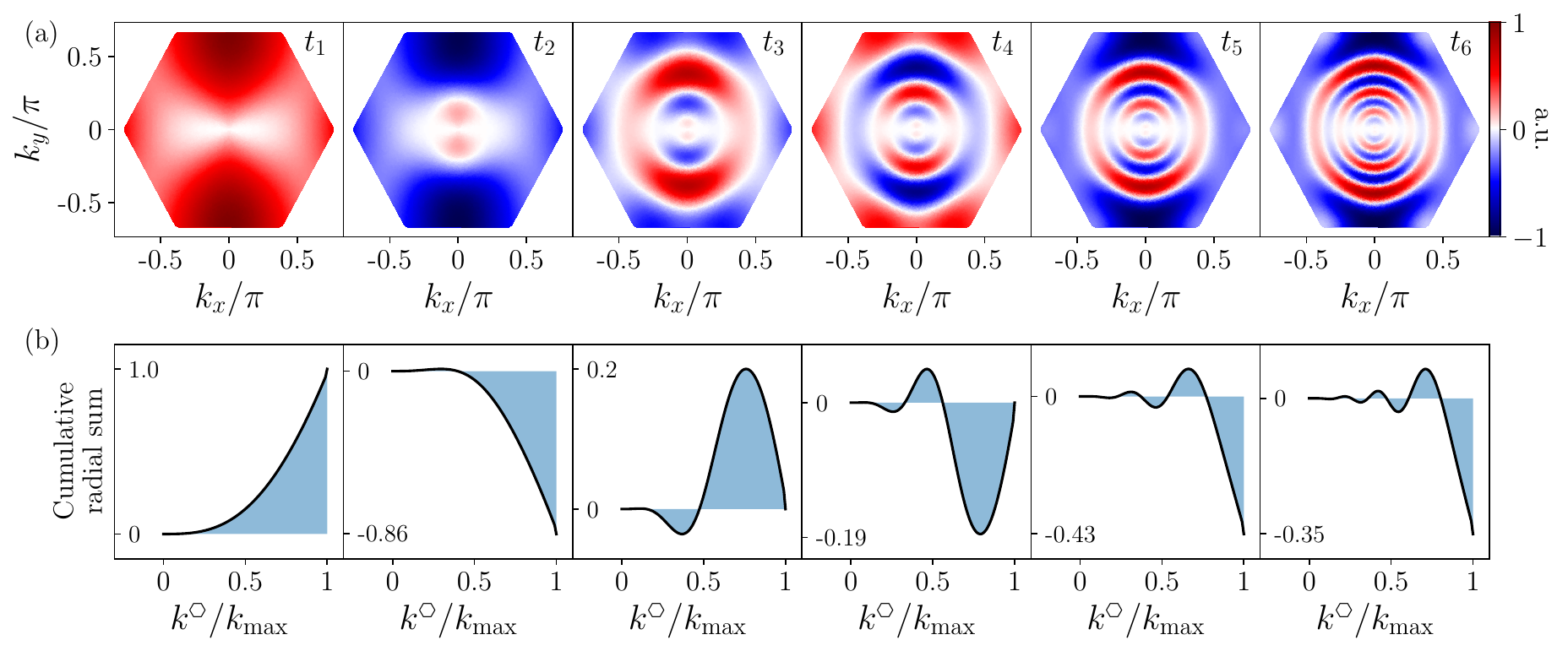}
    \caption{(a) Shows snapshots of $J^y_{\vb*{k}}$ at different times, $t_1 \ldots t_6$, defined in Figure \ref{fig: basic current} with $\theta=0$. In (b) we show the partial sum over $J^y_{\vb*{k}}$ in the plot above. We overlay $J^y_{\vb*{k}}$ with mask with the shape of a smaller hexagon, and increase its size. This means that the final value is equivalent to the total sum over $J^y_{\vb*{k}}$.}
    \label{fig: Jk snapshots}
\end{figure*}

To investigate witch magnons dominate the spin current, we show snapshots of the current distribution  $\vb*{J}_{\vb*{k}}(t)$ in Figure \ref{fig: Jk snapshots} (a), especially at $t_1=0$. Even though the excitations are  distributed over the whole BZ, there are a lot more modes with high frequency. These modes also have a higher amplitude, and thus they dominate. Since not all magnons have the same frequency $\omega_{\vb*{k}}$, concentric rings appear with alternating sign. This dephasing therefore results in cancellation of the contribution to the magnon current, as shown in Figure \ref{fig: Jk snapshots} (b).
Here we plot the cumulative radial sum for each of the snapshots, but instead of concentric circles, we use concentric hexagons with circumradius $k^{\tinyhexagon}$. When the line crosses the $x$-axis at a node $k^{\tinyhexagon}_n$ it means that all contributions from magnons with momentum within $k^{\tinyhexagon}_n$ are cancelled:
\begin{equation}
\sum_{k^{\tinyhexagonn}<k^{\tinyhexagonn}_n} J^{\mu}_{\vb*{k}}=0.
\end{equation}
As time evolves, less and less of the BZ center contributes. Focusing on $t = t_5$, $t_6$, we notice that only magnons with $k^{\tinyhexagon}>k^{\tinyhexagon}_{ln}$, $k^{\tinyhexagon}_{ln}$ being the last node in the cumulative sum, contributes. Therefore, at later times, the amplitude $A(t)$ is reduced as compared to the amplitude directly after the excitation pulse, consistent with the dephasing of magnon modes contributing to the current. This shows that for most of the time, most of the current stems from magnon at the edge of the BZ.

As a function of time, most dephasing occurs near the zone center, which is expected since the magnon dispersion is steeper near the center as compared to the edge. This lets us analyze the decay rate of the two terms in the current in Eq.\eqref{eq: Jk general}. The second term is proportional to $\nabla_{\vb*{k}}\omega_{\vb*{k}}$, and dominates thus more towards the zone center, as compared to the first term proportional to $\nabla_{\vb*{k}}\varphi_{\vb*{k}}$ (See supplementary Figure S.1). Thus, the second term dephases and decays faster.

% , and the term stemming from the broken parity symmetry of the system dominates.

To demonstrate and leverage the coherent nature of the magnon spin current, we perturb the system with different polarized light with a time delay. This creates two currents out of phase in different directions, and results in a total circular current as shown in Figure \ref{fig: tot current, two pulses}. To the best of our knowledge, this effect has not been discussed before.
\begin{figure}[H]
    \centering
    \includegraphics[scale=0.6]{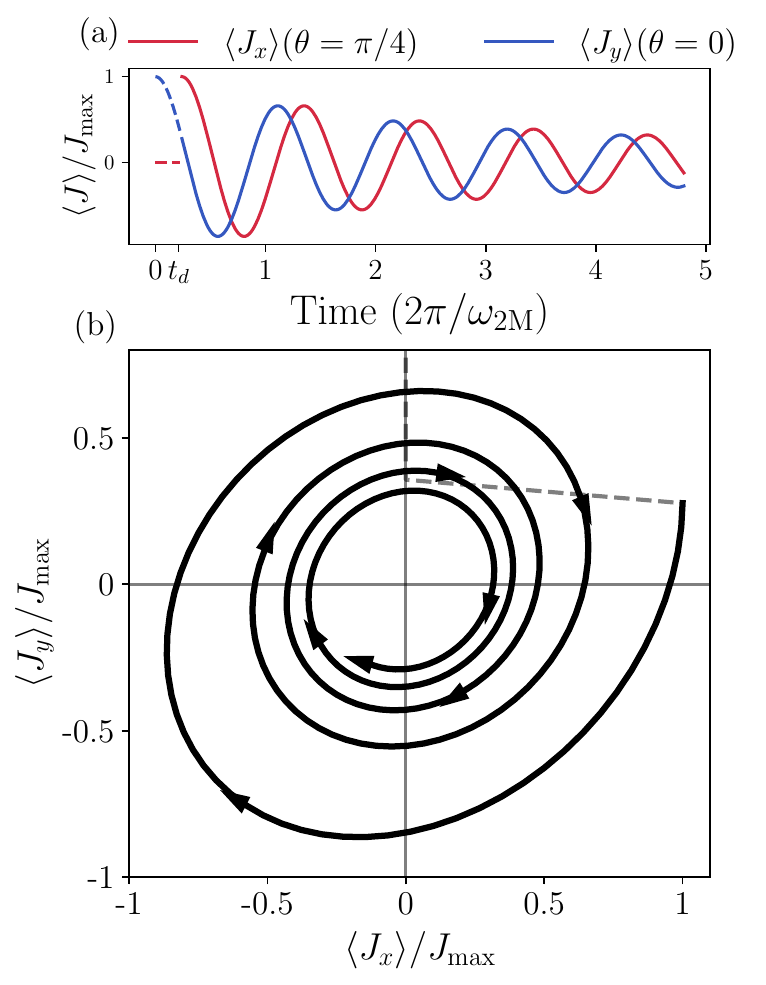}
    \caption{Current as a function of time after two laser pulses. The two quenches are out of phase gives the total current a rotational motion. The dotted lines indicate the evolution of the current after the first pulse, but before the second at $t_d$.}
    \label{fig: tot current, two pulses}
\end{figure}
% \begin{center}
%   \includegraphics[scale=0.6]{figures/circular_current.pdf}
%   \captionof{figure}{Current as a function of time after two laser pulses. The two quenches are out of phase gives the total current a rotational motion. The dotted lines indicate the evolution of the current after the first pulse, but before the second at $t_d$.}
%   \label{fig: tot current, two pulses}
% \end{center}

To estimate the magnitude of the spin current, we focus on the maximum value. In the theoretical model it is at $t=0$, directly after the pulse. Reinserting the lattice constant $a$ and dividing by the thickness $d$ to achieve comparable units to other works \cite{bostromAllopticalGenerationAntiferromagnetic2021}, we get
\begin{equation}
    J_{\mathrm{max}}= |\langle \vb*{J} \rangle_{t=0}| = \frac{3\pi \Delta J S \eta_\tau \Omega}{ad}
\end{equation}
where we introduced the parameter $\eta_\tau = \tau/T_{\mathrm{2M}} \sim 0.01 - 0.1$, which is the ratio between the perturbation strength $\tau$ and the period of the 2M modes at the edge of the BZ, $T_{\mathrm{2M}}$, and $\Omega = 2/N\sum_{\vb*{k}} |\partial_{k_\mu}\gamma_{\vb*{k}}|^2/\epsilon_{\vb*{k}} \approx 0.2$. We use the realistic value $\Delta J \sim 0.01 J_0$ \cite{fedianinSelectionRulesUltrafast2023} and material parameters for \ch{MnPS3}, $S=5/2$, $J_0=1.54$ meV, $a=6.1$Å, and $d=5$Å \cite{bostromAllopticalGenerationAntiferromagnetic2021}. To go from the spin current units of J/$\mathrm{m}^2$ to electrical current density A/$\mathrm{m}^2$, we multiply the current with $e/\hbar$, and  obtain a current amplitude of $J_{\mathrm{max}} \sim 10^{9} - 10^{10}$ A$/\mathrm{m}^2$. 

% \hl{For own sake, check if this is a lot compared to electrical spin currents.}
% \hl{For now, I did not find any number for electron spin currents, but in a wire, this is equivalent to thousands of Ampere..}

The upper estimate is on the same order of magnitude or higher than reported in \cite{bostromAllopticalGenerationAntiferromagnetic2021}, where two magnon modes are also excited, but via circular polarized light under continuous illumination. In works not exciting the two-magnon modes \cite{proskurinExcitationMagnonSpin2018}, the reported current is several orders of magnitude weaker. Another work \cite{fujiwaraNonlinearSpinCurrent2023} reports a magnon shift current from linear polarized light, also several orders of magnitude weaker than our result. We suggest two main reasons for this strong effect. The first is the perturbation of the strong interaction of exchange, and the other is that we excite magnon pairs, utilizing the whole BZ. For experimental detection, with the circular current in Figure \ref{fig: tot current, two pulses}, we except additional circular birefringence on top of the linear birefringence from the two magnons modes.

To detect the two-magnon spin current experimentally, it will be very interesting to investigate the changes of the optical permittivity. In addition to the changes of the diagonal components $\epsilon_{xx}$ and $\epsilon_{yy}$ which are known to be sensitive the two-magnon mode \cite{fedianinSelectionRulesUltrafast2023,formisanoCoherentTHzSpin2024}, the presence of a spin current composed of two-magnon modes will give contributions to the antisymmetric $\epsilon_{xy}\sim \langle \vb*{J} \rangle$, signifying the breaking of time-reversal symmetry. For example, a linear spin current can be measured in a reflection geometry, in which the probing pulse incidents at an oblique angle, while the circular spin current gives a contribution even at normal incidence in a transmission geometry. While current experiments have mostly focused on optical probing \cite{zhaoMagnonSqueezingAntiferromagnet2004,bossiniLaserdrivenQuantumMagnonics2019,formisanoCoherentTHzSpin2024}, it will also be very interesting to study magneto-optical probing in the x-ray regime \cite{laterzaAllopticalSpinInjection2022}.

By modeling a parity-time symmetric Heisenberg antiferromagnet, we have shown that the optical perturbation of exchange interactions can trigger an ultrafast alternating magnon spin current. The frequency of the current and the distribution in reciprocal space gives strong evidence for the current being dominated by magnon pairs close the edge of the BZ. We also show that in the ultrafast regime, coherent control by two orthogonal laser pulses generates a circular magnon-spin current. The main results are obtained by focusing on the prototypical honeycomb lattice that is of key importance for ongoing experiments with van der Waals antiferromagnets \cite{allingtonDistinctOpticalExcitation2025, freemanTunableUltrastrongMagnon2025,kurebayashiMagnetismSymmetrySpin2022,afanasievControllingAnisotropyVan2021,khanRecentBreakthroughsTwodimensional2020,wangProspectsOpportunities2D2020, yangVanWaalsMagnets2021}. We have shown analytically that similar magnon-pair spin currents can be generated if either the lattice itself, the optical perturbations, or both, break parity symmetry, which is further confirmed numerically in supplementary Sec. S.4. Future work may focus on obtaining a deeper understanding of the properties spin current of magnon pairs. For example, magnon-magnon interactions are known to be important at the edge of the BZ which may have profound impact on the current distribution in reciprocal space. Furthermore, it will be very interesting to investigate inhomogeneous systems, such as antiferromagnetic systems featuring domain walls \cite{grundlerNanomagnonicsCorner2016} and study their interaction with the magnon spin current. We hope that our study stimulates experimental investigations of ultrafast nanomagnonics, which might be a very interesting in view of novel femtosecond x-ray spectroscopies.

%\section*{Acknowledgements}
\textbf{Acknowledgement}:
This project received funding from the Dutch Research Council (\href{https://www.nwo.nl/en}{NWO}) \href{https://www.nwo.nl/en/researchprogrammes/nwo-talent-programme/projects-vidi}{VIDI}, nr. 223.157 (\href{https://www.nwo.nl/en/projects/vividi223157}{CHASEMAG}) and from the EU Horizon Europe project nr. \href{https://cordis.europa.eu/project/id/101070290}{101070290} \href{https://www.nimfeia.eu/}{NIMFEIA}.

% This project was funded by the \href{https://www.nimfeia.eu/}{NIMFEIA}: Nonlinear Magnons for Reservoir Computing in Reciprocal Space project of the European Union, under the number \href{https://cordis.europa.eu/project/id/101070290}{101070290}. JHM acknowledges funding by the Dutch Research Council (\href{https://www.nwo.nl/en}{NWO}) via \href{https://www.nwo.nl/en/researchprogrammes/nwo-talent-programme/projects-vidi}{VIDI} project number 223.157 (\href{https://www.nwo.nl/en/projects/vividi223157}{CHASEMAG}).

\end{multicols}

%\bibliographystyle{unsrtnat}
%\bibliography{MyLibrary.bib}

% \twocolumn[
%   \begin{@twocolumnfalse}
%         \printbibliography
%   \end{@twocolumnfalse}
% ]

\vspace{1em}
\noindent\rule{\linewidth}{0.4pt}
\vspace{1em}

\printbibliography[heading=none]

@article{afanasievControllingAnisotropyVan2021,
  title = {Controlling the Anisotropy of a van Der {{Waals}} Antiferromagnet with Light},
  author = {Afanasiev, Dmytro and Hortensius, Jorrit R. and Matthiesen, Mattias and {Ma{\~n}as-Valero}, Samuel and {\v S}i{\v s}kins, Makars and Lee, Martin and Lesne, Edouard and {van der Zant}, Herre S. J. and Steeneken, Peter G. and Ivanov, Boris A. and Coronado, Eugenio and Caviglia, Andrea D.},
  year = {2021},
  month = jun,
  journal = {Science Advances},
  volume = {7},
  number = {23},
  pages = {eabf3096},
  publisher = {American Association for the Advancement of Science},
  doi = {10.1126/sciadv.abf3096},
  urldate = {2025-07-08}
}

@article{allingtonDistinctOpticalExcitation2025,
  title = {Distinct {{Optical Excitation Mechanisms}} of a {{Coherent Magnon}} in a van Der {{Waals Antiferromagnet}}},
  author = {Allington, Clifford J. and Belvin, Carina A. and Seifert, Urban F. P. and Ye, Mengxing and Tai, Tommy and Baldini, Edoardo and Son, Suhan and Kim, Junghyun and Park, Jaena and Park, Je-Geun and Balents, Leon and Gedik, Nuh},
  year = {2025},
  month = feb,
  journal = {Physical Review Letters},
  volume = {134},
  number = {6},
  pages = {066903},
  issn2 = {0031-9007, 1079-7114},
  doi = {10.1103/PhysRevLett.134.066903},
  urldate = {2025-04-04},
  langid = {english}
}

@article{bossiniLaserdrivenQuantumMagnonics2019,
  title = {Laser-Driven Quantum Magnonics and Terahertz Dynamics of the Order Parameter in Antiferromagnets},
  author = {Bossini, D. and Dal Conte, S. and Cerullo, G. and Gomonay, O. and Pisarev, R. V. and Borovsak, M. and Mihailovic, D. and Sinova, J. and Mentink, J. H. and Rasing, {\relax Th}. and Kimel, A. V.},
  year = {2019},
  month = jul,
  journal = {Physical Review B},
  volume = {100},
  number = {2},
  pages = {024428},
  publisher = {American Physical Society},
  doi = {10.1103/PhysRevB.100.024428},
  urldate = {2024-02-09}
}

@article{bossiniMacrospinDynamicsAntiferromagnets2016,
  title = {Macrospin Dynamics in Antiferromagnets Triggered by Sub-20 Femtosecond Injection of Nanomagnons},
  author = {Bossini, D. and Dal Conte, S. and Hashimoto, Y. and Secchi, A. and Pisarev, R. V. and Rasing, Th and Cerullo, G. and Kimel, A. V.},
  year = {2016},
  month = feb,
  journal = {Nature Communications},
  volume = {7},
  number = {1},
  pages = {10645},
  publisher = {Nature Publishing Group},
  issn2 = {2041-1723},
  doi = {10.1038/ncomms10645},
  urldate = {2025-02-17},
  copyright = {2016 The Author(s)},
  langid = {english}
}

@article{bostromAllopticalGenerationAntiferromagnetic2021,
  title = {All-Optical Generation of Antiferromagnetic Magnon Currents via the Magnon Circular Photogalvanic Effect},
  author = {Bostr{\"o}m, Emil Vi{\~n}as and Parvini, Tahereh Sadat and McIver, James W. and Rubio, Angel and Kusminskiy, Silvia Viola and Sentef, Michael A.},
  year = {2021},
  month = sep,
  journal = {Physical Review B},
  volume = {104},
  number = {10},
  pages = {L100404},
  issn2 = {2469-9950, 2469-9969},
  doi = {10.1103/PhysRevB.104.L100404},
  urldate = {2024-05-13},
  langid = {english}
}

@article{boumanTimedependentSchwingerBoson2024,
  title = {Time-Dependent {{Schwinger}} Boson Mean-Field Theory of Supermagnonic Propagation in {{2D}} Antiferromagnets},
  author = {Bouman, M. D. and Mentink, Johan H.},
  year = {2024},
  month = dec,
  journal = {SciPost Physics},
  volume = {17},
  number = {6},
  pages = {159},
  issn2 = {2542-4653},
  doi = {10.21468/SciPostPhys.17.6.159},
  urldate = {2025-01-24},
  langid = {english}
}

@article{chenDampedTopologicalMagnons2023,
  title = {Damped Topological Magnons in Honeycomb Antiferromagnets},
  author = {Chen, Qi-Hui and Huang, Fei-Jie and Fu, Yong-Ping},
  year = {2023},
  month = jul,
  journal = {Physical Review B},
  volume = {108},
  number = {2},
  pages = {024409},
  publisher = {American Physical Society},
  doi = {10.1103/PhysRevB.108.024409},
  urldate = {2025-07-08}
}

@article{chumakMagnonSpintronics2015,
  title = {Magnon Spintronics},
  author = {Chumak, A. V. and Vasyuchka, V. I. and Serga, A. A. and Hillebrands, B.},
  year = {2015},
  month = jun,
  journal = {Nature Physics},
  volume = {11},
  number = {6},
  pages = {453--461},
  publisher = {Nature Publishing Group},
  issn2 = {1745-2481},
  doi = {10.1038/nphys3347},
  urldate = {2025-04-03},
  copyright = {2014 Springer Nature Limited},
  langid = {english}
}

@article{fabianiParametricallyDrivenTHz2022,
  title = {Parametrically Driven {{THz}} Magnon-Pairs: {{Predictions}} toward Ultimately Fast and Minimally Dissipative Switching},
  author = {Fabiani, G. and Mentink, J. H.},
  year = {2022},
  month = apr,
  journal = {Applied Physics Letters},
  volume = {120},
  number = {15},
  publisher = {American Institute of Physics Inc.},
  issn2 = {00036951},
  doi = {10.1063/5.0080161}
}

@article{fabianiSupermagnonicPropagationTwodimensional2021,
  title = {Supermagnonic Propagation in Two-Dimensional Antiferromagnets},
  author = {Fabiani, G. and Bouman, M. D. and Mentink, J. H.},
  year = {2021},
  month = aug,
  journal = {Physical Review Letters},
  volume = {127},
  number = {9},
  eprint = {2101.10945},
  primaryclass = {cond-mat},
  pages = {097202},
  issn2 = {0031-9007, 1079-7114},
  doi = {10.1103/PhysRevLett.127.097202},
  urldate = {2023-10-17},
  archiveprefix = {arXiv}
}

@article{fedianinSelectionRulesUltrafast2023,
  title = {Selection Rules for Ultrafast Laser Excitation and Detection of Spin Correlation Dynamics in a Cubic Antiferromagnet},
  author = {Fedianin, Anatolii E. and Kalashnikova, Alexandra M. and Mentink, Johan H.},
  year = {2023},
  month = apr,
  journal = {Physical Review B},
  volume = {107},
  number = {14},
  pages = {144430},
  publisher = {American Physical Society},
  doi = {10.1103/PhysRevB.107.144430},
  urldate = {2024-06-14}
}

@article{flebus2024MagnonicsRoadmap2024,
  title = {The 2024 Magnonics Roadmap},
  author = {Flebus, Benedetta and Grundler, Dirk and Rana, Bivas and Otani, YoshiChika and Barsukov, Igor and Barman, Anjan and Gubbiotti, Gianluca and Landeros, Pedro and Akerman, Johan and Ebels, Ursula and Pirro, Philipp and Demidov, Vladislav E and Schultheiss, Katrin and Csaba, Gyorgy and Wang, Qi and Ciubotaru, Florin and Nikonov, Dmitri E and Che, Ping and Hertel, Riccardo and Ono, Teruo and Afanasiev, Dmytro and Mentink, Johan and Rasing, Theo and Hillebrands, Burkard and Kusminskiy, Silvia Viola and Zhang, Wei and Du, Chunhui Rita and Finco, Aurore and {van der Sar}, Toeno and Luo, Yunqiu Kelly and Shiota, Yoichi and Sklenar, Joseph and Yu, Tao and Rao, Jinwei},
  year = {2024},
  month = jun,
  journal = {Journal of Physics: Condensed Matter},
  volume = {36},
  number = {36},
  pages = {363501},
  publisher = {IOP Publishing},
  issn2 = {0953-8984},
  doi = {10.1088/1361-648X/ad399c},
  urldate = {2025-06-25},
  langid = {english}
}

@article{flebusRecentAdvancesMagnonics2023,
  title = {Recent Advances in Magnonics},
  author = {Flebus, B. and Rezende, S. M. and Grundler, D. and Barman, A.},
  year = {2023},
  month = apr,
  journal = {Journal of Applied Physics},
  volume = {133},
  number = {16},
  pages = {160401},
  issn2 = {0021-8979},
  doi = {10.1063/5.0153424},
  urldate = {2025-06-26}
}

@article{fleuryScatteringLightOne1968,
  title = {Scattering of {{Light}} by {{One-}} and {{Two-Magnon Excitations}}},
  author = {Fleury, P. A. and Loudon, R.},
  year = {1968},
  month = feb,
  journal = {Physical Review},
  volume = {166},
  number = {2},
  pages = {514--530},
  publisher = {American Physical Society},
  doi = {10.1103/PhysRev.166.514},
  urldate = {2024-08-01}
}

@article{formisanoCoherentTHzSpin2024,
  title = {Coherent {{THz}} Spin Dynamics in Antiferromagnets beyond the Approximation of the {{N{\'e}el}} Vector},
  author = {Formisano, F. and Gareev, T. T. and Khusyainov, D. I. and Fedianin, A. E. and Dubrovin, R. M. and Syrnikov, P. P. and Afanasiev, D. and Pisarev, R. V. and Kalashnikova, A. M. and Mentink, J. H. and Kimel, A. V.},
  year = {2024},
  month = jan,
  journal = {APL Materials},
  volume = {12},
  number = {1},
  pages = {011105},
  issn2 = {2166-532X},
  doi = {10.1063/5.0180888},
  urldate = {2025-01-07}
}

@article{freemanTunableUltrastrongMagnon2025,
  title = {Tunable {{Ultrastrong Magnon}}--{{Magnon Coupling Approaching}} the {{Deep-Strong Regime}} in a van Der {{Waals Antiferromagnet}}},
  author = {Freeman, Charlie W. F. and Youel, Harry and Budniak, Adam K. and Xue, Zekun and De Libero, Henry and Thomson, Thomas and Bosman, Michel and Eda, Goki and Kurebayashi, Hidekazu and Cubukcu, Murat},
  year = {2025},
  month = apr,
  journal = {ACS Nano},
  publisher = {American Chemical Society},
  issn2 = {1936-0851},
  doi = {10.1021/acsnano.5c02576},
  urldate = {2025-04-22}
}

@article{fujiwaraNonlinearSpinCurrent2023,
  title = {Nonlinear Spin Current of Photoexcited Magnons in Collinear Antiferromagnets},
  author = {Fujiwara, Kosuke and Kitamura, Sota and Morimoto, Takahiro},
  year = {2023},
  month = feb,
  journal = {Physical Review B},
  volume = {107},
  number = {6},
  pages = {064403},
  publisher = {American Physical Society},
  doi = {10.1103/PhysRevB.107.064403},
  urldate = {2024-04-26}
}

@article{gillPureSpinCurrents2025,
  title = {Pure {{Spin Currents}} via {{Antisymmetric Light}}},
  author = {Gill, Deepika and Sharma, Sangeeta and Shallcross, Sam},
  year = {2025},
  month = jun,
  journal = {Nano Letters},
  publisher = {American Chemical Society},
  issn2 = {1530-6984},
  doi = {10.1021/acs.nanolett.5c00637},
  urldate = {2025-06-12}
}

@article{gillUltrafastAllopticalGeneration2025,
  title = {Ultrafast All-Optical Generation of Pure Spin and Valley Currents},
  author = {Gill, D. and Sharma, S. and Dewhurst, J. K. and Shallcross, S.},
  year = {2025},
  month = jun,
  journal = {npj 2D Materials and Applications},
  volume = {9},
  number = {1},
  pages = {49},
  publisher = {Nature Publishing Group},
  issn2 = {2397-7132},
  doi = {10.1038/s41699-025-00558-0},
  urldate = {2025-06-26},
  copyright = {2025 The Author(s)},
  langid = {english}
}

@article{goMagnonOrbitalNernst2024,
  title = {Magnon {{Orbital Nernst Effect}} in {{Honeycomb Antiferromagnets}} without {{Spin}}--{{Orbit Coupling}}},
  author = {Go, Gyungchoon and An, Daehyeon and Lee, Hyun-Woo and Kim, Se Kwon},
  year = {2024},
  month = may,
  journal = {Nano Letters},
  volume = {24},
  number = {20},
  pages = {5968--5974},
  publisher = {American Chemical Society},
  issn2 = {1530-6984},
  doi = {10.1021/acs.nanolett.4c00430},
  urldate = {2024-09-13}
}

@article{grundlerNanomagnonicsCorner2016,
  title = {Nanomagnonics around the Corner},
  author = {Grundler, Dirk},
  year = {2016},
  month = may,
  journal = {Nature Nanotechnology},
  volume = {11},
  number = {5},
  pages = {407--408},
  publisher = {Nature Publishing Group},
  issn2 = {1748-3395},
  doi = {10.1038/nnano.2016.16},
  urldate = {2025-06-25},
  copyright = {2016 Springer Nature Limited},
  langid = {english}
}

@article{guptaTuningUltrafastDemagnetization2025,
  title = {Tuning Ultrafast Demagnetization with Ultrashort Spin Polarized Currents in Multi-Sublattice Ferrimagnets},
  author = {Gupta, Deeksha and Pankratova, Maryna and Riepp, Matthias and Pereiro, Manuel and Sanyal, Biplab and Ershadrad, Soheil and Hehn, Michel and Pontius, Niko and {Sch{\"u}{\ss}ler-Langeheine}, Christian and Abrudan, Radu and Bergeard, Nicolas and Bergman, Anders and Eriksson, Olle and Boeglin, Christine},
  year = {2025},
  month = mar,
  journal = {Nature Communications},
  volume = {16},
  number = {1},
  pages = {3097},
  publisher = {Nature Publishing Group},
  issn2 = {2041-1723},
  doi = {10.1038/s41467-025-58411-3},
  urldate = {2025-04-22},
  copyright = {2025 The Author(s)},
  langid = {english}
}

@article{hanCoherentAntiferromagneticSpintronics2023,
  title = {Coherent Antiferromagnetic Spintronics},
  author = {Han, Jiahao and Cheng, Ran and Liu, Luqiao and Ohno, Hideo and Fukami, Shunsuke},
  year = {2023},
  month = jun,
  journal = {Nature Materials},
  volume = {22},
  number = {6},
  pages = {684--695},
  publisher = {Nature Publishing Group},
  issn2 = {1476-4660},
  doi = {10.1038/s41563-023-01492-6},
  urldate = {2025-06-10},
  copyright = {2023 Springer Nature Limited},
  langid = {english}
}

@article{hortensiusCoherentSpinwaveTransport2021,
  title = {Coherent Spin-Wave Transport in an Antiferromagnet},
  author = {Hortensius, J. R. and Afanasiev, D. and Matthiesen, M. and Leenders, R. and Citro, R. and Kimel, A. V. and Mikhaylovskiy, R. V. and Ivanov, B. A. and Caviglia, A. D.},
  year = {2021},
  month = sep,
  journal = {Nature Physics},
  volume = {17},
  number = {9},
  pages = {1001--1006},
  publisher = {Nature Publishing Group},
  issn2 = {1745-2481},
  doi = {10.1038/s41567-021-01290-4},
  urldate = {2025-05-21},
  copyright = {2021 The Author(s), under exclusive licence to Springer Nature Limited},
  langid = {english}
}

@article{ishizukaLargePhotogalvanicSpin2022,
  title = {Large {{Photogalvanic Spin Current}} by {{Magnetic Resonance}} in {{Bilayer Cr Trihalides}}},
  author = {Ishizuka, Hiroaki and Sato, Masahiro},
  year = {2022},
  month = aug,
  journal = {Physical Review Letters},
  volume = {129},
  number = {10},
  pages = {107201},
  issn2 = {0031-9007, 1079-7114},
  doi = {10.1103/PhysRevLett.129.107201},
  urldate = {2025-01-30},
  langid = {english}
}

@article{khanRecentBreakthroughsTwodimensional2020,
  title = {Recent Breakthroughs in Two-Dimensional van Der {{Waals}} Magnetic Materials and Emerging Applications},
  author = {Khan, Yahya and Obaidulla, {\relax Sk}. {\relax Md}. and Habib, Mohammad Rezwan and Gayen, Anabil and Liang, Tao and Wang, Xuefeng and Xu, Mingsheng},
  year = {2020},
  month = oct,
  journal = {Nano Today},
  volume = {34},
  pages = {100902},
  issn2 = {1748-0132},
  doi = {10.1016/j.nantod.2020.100902},
  urldate = {2025-07-08}
}

@article{kurebayashiMagnetismSymmetrySpin2022,
  title = {Magnetism, Symmetry and Spin Transport in van Der {{Waals}} Layered Systems},
  author = {Kurebayashi, Hidekazu and Garcia, Jose H. and Khan, Safe and Sinova, Jairo and Roche, Stephan},
  year = {2022},
  month = mar,
  journal = {Nature Reviews Physics},
  volume = {4},
  number = {3},
  pages = {150--166},
  publisher = {Nature Publishing Group},
  issn2 = {2522-5820},
  doi = {10.1038/s42254-021-00403-5},
  urldate = {2025-07-08},
  copyright = {2022 Springer Nature Limited},
  langid = {english}
}

@article{laterzaAllopticalSpinInjection2022,
  title = {All-Optical Spin Injection in Silicon Investigated by Element-Specific Time-Resolved {{Kerr}} Effect},
  author = {Laterza, Simone and Caretta, Antonio and Bhardwaj, Richa and Flammini, Roberto and Moras, Paolo and Jugovac, Matteo and Rajak, Piu and Islam, Mahabul and Ciancio, Regina and Bonanni, Valentina and Casarin, Barbara and Simoncig, Alberto and Zangrando, Marco and Ribi{\v c}, Primo{\v z} Rebernik and Penco, Giuseppe and Ninno, Giovanni De and Giannessi, Luca and Demidovich, Alexander and Danailov, Miltcho and Parmigiani, Fulvio and Malvestuto, Marco},
  year = {2022},
  month = dec,
  journal = {Optica},
  volume = {9},
  number = {12},
  pages = {1333--1338},
  publisher = {Optica Publishing Group},
  issn2 = {2334-2536},
  doi = {10.1364/OPTICA.471951},
  urldate = {2025-07-21},
  copyright = {{\copyright} 2022 Optica Publishing Group},
  langid = {english}
}

@article{losadaUltrafastControlSpin2019,
  title = {Ultrafast Control of Spin Interactions in Honeycomb Antiferromagnetic Insulators},
  author = {Losada, Juan M. and Brataas, Arne and Qaiumzadeh, Alireza},
  year = {2019},
  month = aug,
  journal = {Physical Review B},
  volume = {100},
  number = {6},
  pages = {060410},
  issn2 = {2469-9950, 2469-9969},
  doi = {10.1103/PhysRevB.100.060410},
  urldate = {2025-06-25},
  langid = {english}
}

@article{maPhotocurrentMultiphysicsDiagnostic2023,
  title = {Photocurrent as a Multiphysics Diagnostic of Quantum Materials},
  author = {Ma, Qiong and Krishna Kumar, Roshan and Xu, Su-Yang and Koppens, Frank H. L. and Song, Justin C. W.},
  year = {2023},
  month = mar,
  journal = {Nature Reviews Physics},
  volume = {5},
  number = {3},
  pages = {170--184},
  publisher = {Nature Publishing Group},
  issn2 = {2522-5820},
  doi = {10.1038/s42254-022-00551-2},
  urldate = {2025-07-08},
  copyright = {2023 Springer Nature Limited},
  langid = {english}
}

@article{mentinkUltrafastReversibleControl2015,
  title = {Ultrafast and Reversible Control of the Exchange Interaction in {{Mott}} Insulators},
  author = {Mentink, J. H. and Balzer, K. and Eckstein, M.},
  year = {2015},
  month = mar,
  journal = {Nature Communications},
  volume = {6},
  number = {1},
  pages = {6708},
  publisher = {Nature Publishing Group},
  issn2 = {2041-1723},
  doi = {10.1038/ncomms7708},
  urldate = {2025-06-24},
  copyright = {2015 The Author(s)},
  langid = {english}
}

@article{pirroAdvancesCoherentMagnonics2021,
  title = {Advances in Coherent Magnonics},
  author = {Pirro, Philipp and Vasyuchka, Vitaliy I. and Serga, Alexander A. and Hillebrands, Burkard},
  year = {2021},
  month = dec,
  journal = {Nature Reviews Materials},
  volume = {6},
  number = {12},
  pages = {1114--1135},
  publisher = {Nature Publishing Group},
  issn2 = {2058-8437},
  doi = {10.1038/s41578-021-00332-w},
  urldate = {2025-06-25},
  copyright = {2021 Springer Nature Limited},
  langid = {english}
}

@article{proskurinExcitationMagnonSpin2018,
  title = {Excitation of Magnon Spin Photocurrents in Antiferromagnetic Insulators},
  author = {Proskurin, Igor and Ovchinnikov, Alexander S. and Kishine, Jun-ichiro and Stamps, Robert L.},
  year = {2018},
  month = oct,
  journal = {Physical Review B},
  volume = {98},
  number = {13},
  pages = {134422},
  issn2 = {2469-9950, 2469-9969},
  doi = {10.1103/PhysRevB.98.134422},
  urldate = {2024-04-02},
  langid = {english}
}

@article{sharmaTHzInducedGiant2023,
  title = {{{THz}} Induced Giant Spin and Valley Currents},
  author = {Sharma, Sangeeta and Elliott, Peter and Shallcross, Samuel},
  year = {2023},
  month = mar,
  journal = {Science Advances},
  volume = {9},
  number = {11},
  publisher = {American Association for the Advancement of Science (AAAS)},
  issn2 = {2375-2548},
  doi = {10.1126/sciadv.adf3673},
  urldate = {2025-07-08},
  langid = {english}
}

@article{wangProspectsOpportunities2D2020,
  title = {Prospects and {{Opportunities}} of {{2D}} van Der {{Waals Magnetic Systems}}},
  author = {Wang, Meng-Chien and Huang, Che-Chun and Cheung, Chi-Ho and Chen, Chih-Yu and Tan, Seng Ghee and Huang, Tsung-Wei and Zhao, Yue and Zhao, Yanfeng and Wu, Gang and Feng, Yuan-Ping and Wu, Han-Chun and Chang, Ching-Ray},
  year = {2020},
  journal = {Annalen der Physik},
  volume = {532},
  number = {5},
  pages = {1900452},
  issn2 = {1521-3889},
  doi = {10.1002/andp.201900452},
  urldate = {2025-07-08},
  copyright = {{\copyright} 2020 The Authors. Published by WILEY-VCH Verlag GmbH \& Co. KGaA, Weinheim},
  langid = {english}
}

@article{yangVanWaalsMagnets2021,
  title = {Van Der {{Waals Magnets}}: {{Material Family}}, {{Detection}} and {{Modulation}} of {{Magnetism}}, and {{Perspective}} in {{Spintronics}}},
  shorttitle = {Van Der {{Waals Magnets}}},
  author = {Yang, Shengxue and Zhang, Tianle and Jiang, Chengbao},
  year = {2021},
  journal = {Advanced Science},
  volume = {8},
  number = {2},
  pages = {2002488},
  issn2 = {2198-3844},
  doi = {10.1002/advs.202002488},
  urldate = {2025-07-08},
  copyright = {{\copyright} 2020 The Authors. Advanced Science published by Wiley-VCH GmbH},
  langid = {english}
}

@article{zeleznySpinTransportSpin2018,
  title = {Spin Transport and Spin Torque in Antiferromagnetic Devices},
  author = {{\v Z}elezn{\'y}, J. and Wadley, P. and Olejn{\'i}k, K. and Hoffmann, A. and Ohno, H.},
  year = {2018},
  month = mar,
  journal = {Nature Physics},
  volume = {14},
  number = {3},
  pages = {220--228},
  publisher = {Nature Publishing Group},
  issn2 = {1745-2481},
  doi = {10.1038/s41567-018-0062-7},
  urldate = {2025-07-08},
  copyright = {2018 Springer Nature Limited},
  langid = {english}
}

@article{zhaoMagnonSqueezingAntiferromagnet2004,
  title = {Magnon {{Squeezing}} in an {{Antiferromagnet}}: {{Reducing}} the {{Spin Noise}} below the {{Standard Quantum Limit}}},
  shorttitle = {Magnon {{Squeezing}} in an {{Antiferromagnet}}},
  author = {Zhao, Jimin and Bragas, A. V. and Lockwood, D. J. and Merlin, R.},
  year = {2004},
  month = sep,
  journal = {Physical Review Letters},
  volume = {93},
  number = {10},
  pages = {107203},
  publisher = {American Physical Society},
  doi = {10.1103/PhysRevLett.93.107203},
  urldate = {2025-02-17}
}

\vspace{1em}
\noindent\rule{\linewidth}{0.4pt}
\vspace{1em}

\section*{Supplementary: Ultrafast coherent magnon spin currents in antiferromagnets}

\renewcommand{\theequation}{S.\arabic{equation}}
\renewcommand{\thefigure}{S.\arabic{figure}}
\renewcommand{\thesection}{S.\arabic{section}}

% reset the counter
\setcounter{equation}{0}
\setcounter{figure}{0}

\section{Time dependence calculation}
Staying within linear response, we can ignore the effect of $\alpha_{\vb*{k}}^\dag \alpha_{\vb*{k}}$ and $\beta_{-\vb*{k}}^\dag \beta_{-\vb*{k}}$ as done in \cite{boumanTimedependentSchwingerBoson2024}. The expectation value of some operator $A$, is then given by
\begin{equation}
    \begin{split}
    \langle A \rangle(t) =  \langle A \rangle_0 &- i \sum_{\vb*{k}} \left\langle \left[ A_I(t), \alpha_{\vb*{k}} \beta_{\vb*{k}} \right] \right\rangle_0 V^*_{\vb*{k}}\int_{-\infty}^t \dd t' f(t') e^{-i2\omega_{\vb*{k}} t'/\hbar}\\
    &-i \sum_{\vb*{k}} \left\langle \left[ A_I(t), \alpha_{\vb*{k}}^\dag \beta_{\vb*{k}}^\dag \right] \right\rangle_0 V_{\vb*{k}}\int_{-\infty}^t \dd t' f(t') e^{i2\omega_{\vb*{k}} t'/\hbar}
\end{split}
\end{equation}
where $A_I = e^{iH_0t} A e^{-iH_0 t}$ is the operator in interaction picture. Using $f(t)=\tau\delta(t)$, we get
\begin{equation}
    \langle \alpha_{\vb*{k}}^\dag \beta_{-\vb*{k}}^\dag  \rangle = - i \sum_{\vb*{k}'} -e^{2i\omega_{\vb*{k}'}t} \delta_{\vb*{k}, \vb*{k}'} V_{\vb*{k}'}^* \tau = V_{\vb*{k}}^*\tau i e^{2i\omega_{\vb*{k}}t}
\end{equation}

\section{Parity symmetry}
\label{sect: Inversion symmetry}
To simplify some of the expressions in this work, we will look into the symmetries of the relevant functions. We start with
\begin{equation}
    \mathrm{Re}\{\gamma_{\vb*{k}} \} = \frac{1}{z}\sum_{\vb*{\delta}} \rho_{\vb*{\delta}}\cos({\vb*{k}} \cdot \vb*{\delta}), \qquad \mathrm{Im}\{\gamma_{\vb*{k}} \} = \frac{1}{z}\sum_{\vb*{\delta}} \rho_{\vb*{\delta}}\sin({\vb*{k}}\cdot \vb*{\delta})
\end{equation}
where the exchange in direction $\vb*{\delta}$, $J_{\vb*{\delta}} = J_0 \rho_{\vb*{\delta}}$. For the honeycomb lattice, $\rho_{\vb*{\delta}}=1$.
From this it is easy to see that $\mathrm{Re}\{\gamma_{\vb*{k}} \}$ is even and $\mathrm{Im}\{\gamma_{\vb*{k}} \} $ is odd in $\vb*{k}$. With this, we define
\begin{equation}
    |\gamma_{\vb*{k}}| = \sqrt{\mathrm{Re}\{\gamma_{\vb*{k}} \}^2 + \mathrm{Im}\{\gamma_{\vb*{k}} \}^2}, \qquad \varphi_{\vb*{k}} = \arctan{\frac{\mathrm{Im}\{\gamma_{\vb*{k}} \}}{\mathrm{Re}\{\gamma_{\vb*{k}} \}}}
\end{equation}
and see that $|\gamma_{\vb*{k}}|$ is even and $\varphi_{\vb*{k}}$ is odd. A similar argument holds for $\xi_{\vb*{k}}$; $\mathrm{Re}\{\xi_{\vb*{k}}\}$ is even while $\mathrm{Im}\{\xi_{\vb*{k}} \}$ is odd. From this one can easily show that $\mathrm{Re}\{V_{\vb*{k}}\}$ is even and $\mathrm{Im}\{V_{\vb*{k}}\}$ is odd. Also, $\epsilon_{\vb*{k}}=\sqrt{1-|\gamma_{\vb*{k}}|^2}$ is even.

\section{Expression for the spin current}
The expression for the current can be written as
\begin{equation}
    \label{eq: full current expression}
    \begin{split}
    \langle \vb*{J} \rangle = \frac{4}{N}\sum_{\vb*{k}}\mathrm{Re}\{\vb*{K}_{\vb*{k}}\langle \alpha_{\vb*{k}}^\dagger \beta_{-\vb*{k}}^\dag \rangle  \} = \frac{4zJ_0S\tau}{N}\sum_{\vb*{k}}\bigg[&
    \bigg( |\gamma_{\vb*{k}}| \nabla_{\vb*{k}} \varphi_{\vb*{k}} \cos{\left(2 \omega_{\vb*{k}} t \right)} - \frac{\nabla_{\vb*{k}} |\gamma_{\vb*{k}}| \sin{\left(2 \omega_{\vb*{k}} t \right)}}{\epsilon_{\vb*{k}}} \bigg) \mathrm{Re}\{V_{\vb*{k}}\}\\
    - &\bigg(|\gamma_{\vb*{k}}| \nabla_{\vb*{k}} \varphi_{\vb*{k}} \sin{\left(2 \omega_{\vb*{k}} t \right)} + \frac{\nabla_{\vb*{k}} |\gamma_{\vb*{k}}| \cos{\left(2 \omega_{\vb*{k}} t \right)}}{\epsilon_{\vb*{k}}} \bigg) \mathrm{Im}\{V_{\vb*{k}}\}\bigg]
    \end{split}
\end{equation}
Now, based in the results in section \ref{sect: Inversion symmetry}, we can see which terms are odd in $\vb*{k}$. The second term consists of the odd factor $\nabla_{\vb*{k}}\gamma_{\vb*{k}}$ and the even factors $\sin(2\omega_{\vb*{k}}t)$, $\mathrm{Re}\{V_{\vb*{k}}\}$, and $\epsilon_{\vb*{k}}$ making the whole term odd, which vanishes after summation. The same conclusion can be drawn for the third term, giving us Eq. (10) in the main paper.

\subsection{Distribution of the spin current}
\begin{figure}[H]
    \centering
    \includegraphics[scale=0.7]{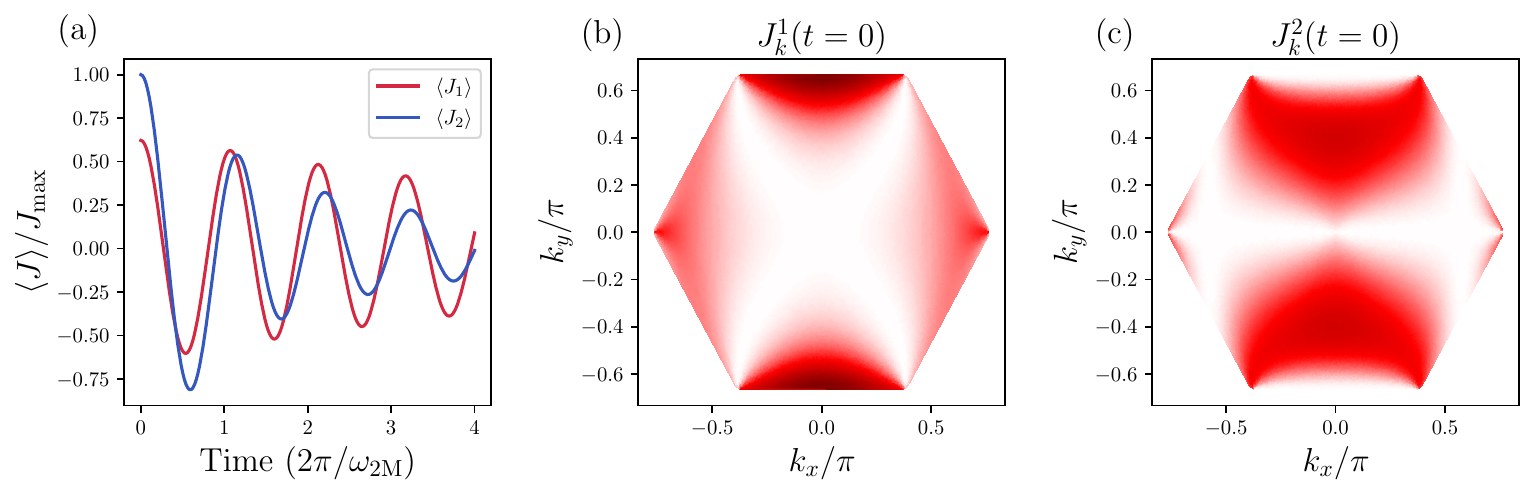}
    \caption{(a) shows the time dependence of the two current contributions $\langle J_1 \rangle$ and $\langle J_2 \rangle$ in the $y$-direction with polarization angle $\theta=0$. (b)[(c)] shows the current contribution distribution over the BZ for $\langle J_1 \rangle$ $[\langle J_2 \rangle]$}.
    \label{fig: current terms distr}
\end{figure}

The two terms after application of parity symmetry analysis of Eq. \eqref{eq: full current expression} take the forms
\begin{equation}
    \vb*{J}_{\vb*{k}}^1  \sim   |\gamma_{\vb*{k}}| \nabla_{\vb*{k}} \varphi_{\vb*{k}} \mathrm{Re}\{V_{\vb*{k}}\} \cos(2\omega_{\vb*{k}}t),
\end{equation}
\begin{equation}
    \vb*{J}_{\vb*{k}}^2 \sim \frac{\nabla_{\vb*{k}}\epsilon_{\vb*{k}}}{|\gamma_{\vb*{k}}|} \mathrm{Im}\{V_{\vb*{k}}\} \cos(2\omega_{\vb*{k}}t)
\end{equation}
where we have used the relation $\nabla_{\vb*{k}} \epsilon_{\vb*{k}} = -|\gamma_{\vb*{k}}| \nabla_{\vb*{k}} |\gamma_{\vb*{k}}|/\epsilon_{\vb*{k}}$. In Figure \ref{fig: current terms distr} we show how the $y$-component ($\theta=0$) of the two terms evolve over time; with different decay and period. The figure also shows the current contribution distribution at $t=0$, revealing that $\langle J_1 \rangle$ mostly consists of modes close to the edge of the BZ, while $\langle J_2 \rangle$ is more distributed closer to the center of the BZ.

\section{Square lattice}

\begin{figure}[H]
    \centering
    \includegraphics[scale=0.8]{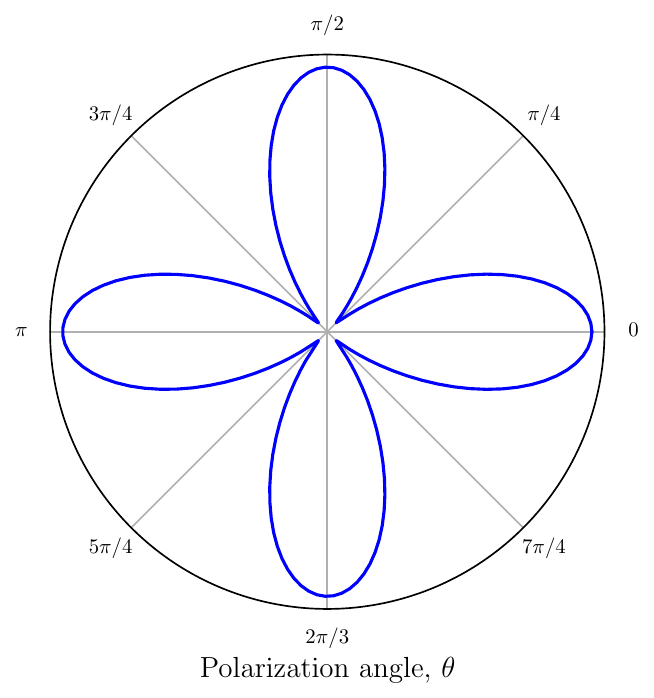}
    \caption{Polar plot of the current amplitude as a function of polariation angle for a square lattice with party-broken exchange.}
    \label{fig: polarization dep broken square}   
\end{figure}

In this section we show that a magnon spin current is also possible to achieve in a square lattice with a parity broken exchange Heisenberg Hamiltonian given by
\begin{eqnarray}
    \mathcal{H}_0 = J_0 \sum_{i,\vb*{\delta}} \rho_{\vb*{\delta}} \vb*{S}_i \cdot \vb*{S}_{i+\delta}
\end{eqnarray}
and $\delta \mathcal{H}$ is the same as in the main work. Using the same procedure as in the main work, with  $\{ \rho_{\vb*{\delta}} \}=\begin{bmatrix} \rho_{-\vb*{x}} & \rho_{\vb*{x}} & \rho_{-\vb*{y}} & \rho_{\vb*{y}} \end{bmatrix}= \begin{bmatrix}0.9& 1.1 & 0.9 & 1.1\end{bmatrix}$, we get a current with polarization dependence given in Figure \ref{fig: polarization dep broken square}.

Introducing a similar $\rho_{\vb*{\delta}}$ to $\delta\mathcal{H}$ will also give a spin current. Thus, it is enough to break parity symmetry in either $\mathcal{H}_0$ or $\delta\mathcal{H}$.

\end{document}